\documentclass[twocolumn,prl,showpacs,superscriptaddress,showpacs]{revtex4}
\usepackage{amssymb,amsmath}
\usepackage{graphicx}
\usepackage{bm}
\usepackage{dcolumn}
\usepackage{float}
\usepackage{url}
\usepackage{color}
\usepackage{subfigure}
\usepackage{siunitx}
\usepackage{txfontsb}
\usepackage[OT1]{fontenc}
\usepackage[driverfallback=dvipdfm]{hyperref}
\usepackage{soul,ulem}
\hypersetup{colorlinks=true,breaklinks,urlcolor=blue,linkcolor=blue,citecolor=blue}

\begin{document}

\title{Measuring the Chern-Simons invariant in quantum gases}


\author{Chang-Rui Yi}
\thanks{These authors contribute equally to this work.}
\affiliation{Hefei National Research Center for Physical Sciences at the Microscale and School of Physical Sciences, University of Science and Technology of China, Hefei 230026, China}
\affiliation{Shanghai Research Center for Quantum Science and CAS Center for Excellence in Quantum Information and Quantum Physics, University of Science and Technology of China, Shanghai 201315, China}
\affiliation{Hefei National Laboratory, University of Science and Technology of China, Hefei 230088, China
}

\author{Jinlong Yu}
\thanks{These authors contribute equally to this work.}
\affiliation{Center for Theoretical Physics, Hainan University, Haikou 570228, China}
\affiliation{School of Physics and Optoelectronic Engineering, Hainan University, Haikou 570228, China
}

\author{Huan Yuan}
\thanks{These authors contribute equally to this work.}
\affiliation{Hefei National Research Center for Physical Sciences at the Microscale and School of Physical Sciences, University of Science and Technology of China, Hefei 230026, China}
\affiliation{Shanghai Research Center for Quantum Science and CAS Center for Excellence in Quantum Information and Quantum Physics, University of Science and Technology of China, Shanghai 201315, China}
\affiliation{Hefei National Laboratory, University of Science and Technology of China, Hefei 230088, China
}

\author{Xin Chen}
\affiliation{Institute for Theoretical Physics, Heidelberg University, Philosophenweg 16, 69120 Heidelberg, Germany}

\author{Jia-Yu Guo}
\affiliation{Hefei National Research Center for Physical Sciences at the Microscale and School of Physical Sciences, University of Science and Technology of China, Hefei 230026, China}
\affiliation{Shanghai Research Center for Quantum Science and CAS Center for Excellence in Quantum Information and Quantum Physics, University of Science and Technology of China, Shanghai 201315, China}
\affiliation{Hefei National Laboratory, University of Science and Technology of China, Hefei 230088, China
}

\author{Jinyi Zhang}
\affiliation{Hefei National Research Center for Physical Sciences at the Microscale and School of Physical Sciences, University of Science and Technology of China, Hefei 230026, China}
\affiliation{Shanghai Research Center for Quantum Science and CAS Center for Excellence in Quantum Information and Quantum Physics, University of Science and Technology of China, Shanghai 201315, China}
\affiliation{Hefei National Laboratory, University of Science and Technology of China, Hefei 230088, China
}

\author{Shuai Chen}
\affiliation{Hefei National Research Center for Physical Sciences at the Microscale and School of Physical Sciences, University of Science and Technology of China, Hefei 230026, China}
\affiliation{Shanghai Research Center for Quantum Science and CAS Center for Excellence in Quantum Information and Quantum Physics, University of Science and Technology of China, Shanghai 201315, China}
\affiliation{Hefei National Laboratory, University of Science and Technology of China, Hefei 230088, China
}

\author{Jian-Wei Pan}
\affiliation{Hefei National Research Center for Physical Sciences at the Microscale and School of Physical Sciences, University of Science and Technology of China, Hefei 230026, China}
\affiliation{Shanghai Research Center for Quantum Science and CAS Center for Excellence in Quantum Information and Quantum Physics, University of Science and Technology of China, Shanghai 201315, China}
\affiliation{Hefei National Laboratory, University of Science and Technology of China, Hefei 230088, China
}

\date{\today}

\begin{abstract}
Chern-Simons (CS) invariant is a fundamental topological invariant describing the topological invariance of 3D space based on the Chern-Simons field theory.
To date, direct measurement of the CS invariant in a physical system remains elusive. 
Here, the CS invariant is experimentally measured by quenching a 2D optical Raman lattice with 1/2 spin in ultracold atoms.
With a recently developed Bloch state tomography, we measure the expectation values of three Pauli matrices in 2D quasi-momentum space plus 1D time [(2+1)D], and then respectively extract the Berry curvature and the corresponding Berry connection.
By integrating the product of these two quantities, we obtain the CS invariants near $\pm 1$ and 0, which are consistent with theoretical predictions.
We also observe transitions among these values, which indicates the change of the topology of the quantum state in (2+1)D quantum dynamics.
\end{abstract}

\maketitle
In 1974, Chern and Simons studied the geometrical properties of smooth 3D space and advanced a topological invariant that describes the topological invariance of 3D space, i.e., Chern-Simons invariant \cite{Chern-Simons}.
In the late 1980s, such geometrical perspectives were introduced to the quantum field theory.
Witten established the Chern-Simons field theory and pointed out that many topological invariants of the knots and links discovered in the knot theory can be reinterpreted as the correlation functions of Wilson loop operators \cite{witten1988,witten1989}.
Meanwhile, in condensed matter physics, Chern-Simons field theory was also developed to elaborate that the topological insulators and the fractional quantum Hall effect are dominated by the low-energy effective actions \cite{topo_insu_super,PhysRevLett.62.82}.
In particular, the CS invariant calculated by the low-energy effective theory is an essential physical quantity for exhibiting the classification of 3D topological materials \cite{PhysRevB.78.195424,Chiu2016}.

\begin{figure*}
\begin{center}
\includegraphics[width=1\linewidth]{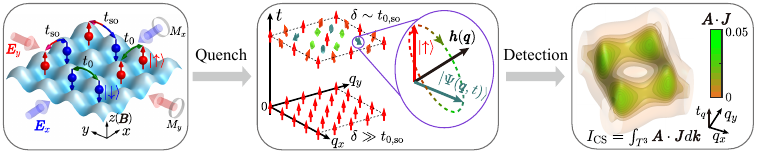}
\caption{The scheme for measuring the CS invariant in a quenched Raman lattice.
Left: cartoon of the Raman lattices.
The Raman lattices are realized by two reflected beams $E_{x,y}$.
$M_{x,y}$ are Mirrors.
Middle: scheme of the quench.
The initial states are polarized.
After suddenly switching detuning $\delta$ to $\delta\sim t_{0,\rm{so}}$, the evolutive states precess surrounding the vector $\bm{h}(\bm{q})$.
Right: obtaining the CS invariant.
The CS invariant $I_{\rm{CS}}$ is determined by Eq. (\ref{CSinvarint}).
}
\label{Fig1}
\end{center}
\end{figure*}

For a closed 3D space $\mathcal{M}$, the Abelian CS invariant $I_{\rm{CS}}$\cite{note3} is defined as the integral of the product of the Berry curvature $\bm{J}(\bm{k})$ and the Berry connection $\bm{A}(\bm{k})$ over all $\bm{k}=(k^x,k^y,k^z)\in \mathcal{M}$ points~\cite{Chern-Simons,PhysRevA.64.052101,linking_HuiZhai,Wilczek_Zee1983,WU1984325}, i.e.,
\begin{equation}\label{CSinvarint}
I_{\rm{CS}}=\int_{\mathcal{M}} \bm{A}(\bm{k})\cdot \bm{J}(\bm{k})d\bm{k},
\end{equation}
where the component of the Berry connection is $A_{\mu}(\bm{k})=i\langle \Psi(\bm{k})|\partial_{\mu}| \Psi(\bm{k})\rangle/(2\pi)$ with the quantum state $| \Psi(\bm{k})\rangle$ and $\partial_{\mu}=\partial/\partial {k^{\mu}}$ ($\mu=x,y,z$); the Berry curvature is $\bm{J}(\bm{k})=\nabla\times \bm{A}(\bm{k})$.
The CS invariant has been widely investigated in a variety of theoretical domains \cite{Guadagnini1993,Jorgen2011,Chiu2016,RevModPhys.77.675}.
However, experiments for directly measuring the CS invariant as integration of Eq.(\ref{CSinvarint}) remain to be elusive since
the direct measurement of 3D distribution of Berry curvature is a serious challenge~\cite{note2}.

Recently, in ultracold atoms, Ref.~\cite{linking_HuiZhai} reveals a topological invariant in (2+1)D based on quench dynamics of a two-band topological system with Hamiltonian $H$, the definition of which is consistent with Eq.(\ref{CSinvarint}).
Here, the quantum state is written as $|\Psi(\bm{q},t)\rangle=e^{-iHt}|\Psi_i\rangle$ (the reduced Planck constant $\hbar=1$), where $\bm{q}=(q_{x},q_y)$, $t$ and $|\Psi_i\rangle$ are respectively the quasi-momentum, time of the quench and the initial quantum state.
With the perception of Ref.~\cite{PhysRevA.64.052101}, such invariant in Ref.~\cite{linking_HuiZhai} can be regarded as the CS invariant.
The key points for implementing the proposal of Ref.~\cite{linking_HuiZhai} are the following:
i) a well-controlled 2D topological system should be constructed.
ii) a sufficiently long coherence time should be maintained during the quench dynamics.
iii) complete Bloch state tomography are required to be performed, which enables directly extract the 3D distribution of the Berry curvature.
In current experiments, quench dynamics has been used to identify the topology of a shaken hexagonal optical lattice \cite{dynamicalVortices_Hamburg2017,link2019} and a 2D optical Raman lattice \cite{uncover_topology,Yi2019}.
Moreover, the Bloch state tomography has also been developed in those systems~\cite{sengstock_Berry_curvature,yi2023extracting}.
Finally, the current techniques of ultracold atoms render a well-controlled testbed for probing the CS invariant.

In this work, we implement the measurements of the CS invariant in a quenched 2D optical Raman lattice~\cite{realization2DSOC,W.S_longlive} for ultracold atoms by improving the current techniques.
Thanks to the Bloch state tomography in the Raman lattices \cite{yi2023extracting}, the distributions of the expectation values of three Pauli matrices are observed in (2+1)D, with which the vectors of the Berry curvature and the corresponding Berry connection under a particular gauge are extracted.
Using the integral of the product of these two vectors, the CS invariant is determined.
Further, the CS invariant valued around $\pm1$ and 0 are observed via altering a parameter of the Raman lattices.
The transitions among these values reveal the change of topology of the quantum state in (2+1)D.

The scheme for measuring the CS invariant is realized by the quench dynamics of a 2D quantum anomalous Hall (QAH) model, as depicted in the left of Fig.~\ref{Fig1}.
This model is demonstrated in an optical Raman lattice with ultracold ${}^{87}$Rb atoms constructed by two orthogonal beams $\bm{E}_{x,y}$ ~\cite{supMat,realization2DSOC,W.S_longlive,Realization2DSOCtheory}.
Two Raman processes generated by these two beams couple the magnetic sublevels $|F=1, m_F=-1\rangle$ (spin up $\mid\uparrow\rangle$) and $|F=1, m_F=0\rangle$ (spin down $\mid\downarrow\rangle$), being split by the bias magnetic field $\bm{B}$ with a magnitude of 23.4G.
The Hamiltonian of the QAH model reads $\mathcal{H}(\bm{q})=\bm{h}(\bm{q})\cdot \bm{\sigma}$,
where the vector $\bm{h}(\bm{q})=(h_x, h_y, h_z)$ with $h_x=2t_{\rm{so}}\sin{q_y}$, $h_y=2t_{\rm{so}}\sin{q_x}$, $h_z=\delta/2-2t_0(\cos{q_x}+\cos{q_y})$, and the Pauli matrices $\bm{\sigma}=(\sigma_x,\sigma_y,\sigma_z)$; we set the lattice constant as the length unit.
Here, $\delta$ is the two-photon detuning and $t_{\rm{0}}$ ($t_{\rm{so}}$) is the spin-conserved (spin-flipped) hopping coefficient.
This model possesses topologically non-trivial (trivial) regime with $0<|\delta|<8t_0$ ($|\delta|>8t_0$) \cite{supMat}.

The quench process is achieved by suddenly changing the Hamiltonian $\mathcal{H}(\bm{q})$ from a topologically trivial Hamiltonian to a final Hamiltonian [the middle of Fig.~\ref{Fig1}].
In particular, we prepare the initial trivial state as a polarized one $|\Psi(\bm{q},t=0)\rangle\equiv\mid\uparrow\rangle$ by setting $\delta\gg t_{0,\rm{so}}$.
After suddenly switching the detuning to $\delta\sim t_{0,\rm{so}}$, the initial state unitarily evolves surrounding the vector $\bm{h}(\bm{q})$ and then the evolved state is written as $\left| {\Psi ({\bm{q}},t)} \right\rangle  = \exp [ - i\mathcal{H}(\bm{q}) t]\mid\uparrow\rangle$.
Thus, the expectation values of three Pauli matrices ${\bm{P}}({\bm{q}},t) = \left\langle {\Psi ({\bm{q}},t)} \right|\bm{\sigma} \left| {\Psi ({\bm{q}},t)} \right\rangle$ is extracted by the projected measurements of the evolved state, i.e.
\begin{equation} \label{Eq:Hopf_map_T3_S2}
	\begin{aligned}
  {P_x}({\bm{q}},t_q) &= \sin (t_q){{\hat h}_y} + [1 - \cos (t_q)]{{\hat h}_x}{{\hat h}_z}, \hfill \\
  {P_y}({\bm{q}},t_q) &=  - \sin (t_q){{\hat h}_x} + [1 - \cos (t_q)]{{\hat h}_y}{{\hat h}_z}, \hfill \\
  {P_z}({\bm{q}},t_q) &= \cos (t_q) + [1 - \cos (t_q)]\hat h_z^2, \hfill \\
\end{aligned}
\end{equation}
where the normalized vector $\bm{\hat h}(\bm{q}) = {(\hat h_x, \hat h_y, \hat h_z)} = -\bm{h}(\bm{q})/|\bm{h}(\bm{q})|$.
Here, we have defined a rescaled time $t_q=2|\bm{h}(\bm{q})|t$, which features a period of $2\pi$ just as the quasi-momenta $q_x$ or $q_y$ do.
Accordingly, combining $\bm{q}$ and $t_q$, the 3D torus $T^3=T^2\times S^1$ is constructed, where the 2D torus $T^2$ stands for the lattice Brillouin zone with $q_{x,y}\in[-\pi,\pi)$ and the rescaled time forms a circle $t_q \in [0,2\pi )\in S^1$.

Based on $\bm{P}(\bm{k})$ with $\bm{k}=(q_x,q_y,t_q)$, the CS invariant $I_{\rm{CS}}$ is determined from Eq. (\ref{CSinvarint})
[the right of Fig.~\ref{Fig1}].
Each components of the Berry curvature take the form ${J}_{\mu}(\bm{k})=\epsilon_{\mu \nu \lambda}\bm{P}(\bm{k})\cdot \left [\partial_{\nu}\bm{P}(\bm{k}) \times \partial_{\lambda}\bm{P}(\bm{k})\right ]/(8\pi)$ ~\cite{Wilczek_Zee1983,WU1984325,linking_HuiZhai}.
The indexes $\mu$, $\nu$ and $\lambda$ take values in $q_x$, $q_y$ and $t_q$, as well as $\epsilon_{\mu \nu \lambda}$ is the totally antisymmetric tensor that is adopted throughout the Einstein's summation convention.
Each components of $\bm{A}(\bm{k})$ can be extracted from $J_{\mu}(\bm{k})$, namely ${A}_{\mu}(\bm{k})=\int_{T^3} {J}_{\mu}(\bm{k}^{\prime})\times (\bm{k}-\bm{k}^{\prime})/|\bm{k}-\bm{k}^{\prime}|^3d\bm{k}^{\prime}/(4\pi)$ with the gauge $\partial_{\mu}A_{\mu}=0$~\cite{Hopflinks}.
Thus, we obtain three quantized values of the CS invariant: $I_{\rm{CS}}=+1~(-1)$ for $0<\delta<8t_0~(-8t_0<\delta<0)$; $I_{\rm{CS}}=0$ for $|\delta|>8t_0$.

\begin{figure*}[t]
\begin{center}
\includegraphics[width=0.8\linewidth]{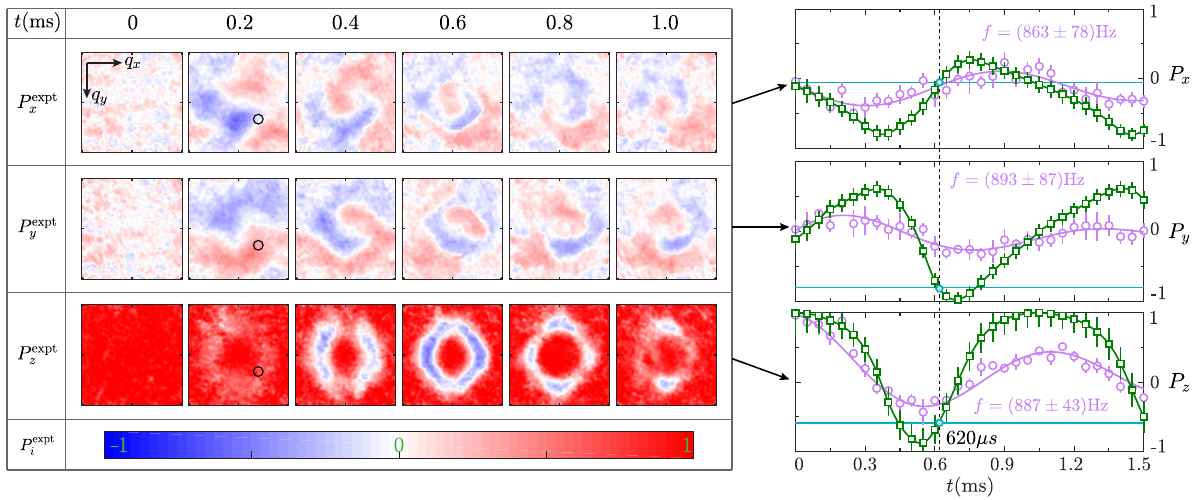}
\caption{The expectation values of three Pauli matrices $P_{x,y,z}(\bm{q},t)$ with $\delta=0.20(3)E_{\rm{r}}$.
Left: measurements of time evolution of $P_{x,y,z}^{\rm{exp}}(\bm{q},t)$ in the lattice Brillouin zone.
Right: time evolution of $P_{x,y,z}^{\rm{exp}}(\bm{q},t)$ at quasi-momentum point $\bm{q}=(0.39,0.33)\pi$, whose location is marked by circles in the $P_{x,y,z}^{\rm{exp}}(\bm{q},t=0.2\text{ms})$.
The purple circles with error bars are from original experimental measurements.
The green squares with error bars are from the components of the normalized $P_{x,y,z}^{\rm{exp}}(\bm{q},t)$.
The purple solid curves is the fittings from sinusoidal function with damping~\cite{supMat}.
The cyan straight lines from top to bottom are respectively obtained by $\sin\theta\cos\varphi$, $\sin\theta\sin\varphi$ and $\cos\theta$ with $\theta=0.7\pi$ and $\varphi=1.48\pi$.
The intersection of the cyan straight lines and green curves give the location of a point on a closed loop.
}
\label{Fig2}
\end{center}
\end{figure*}

In the experiment, we perform the quench dynamics in the 2D Raman lattices~\cite{uncover_topology,windingNumber_zhang} to determine the CS invariant by detecting the expectation values of three Pauli matrices in $\bm{k}$ space.
The experimental protocol is performed as follows \cite{supMat}.
Firstly, the ${}^{87}\textrm{Rb}$ atoms are prepared slightly above the critical temperature of Bose-Einstein condensation and then loaded into each momentum point in the lowest band of the Raman lattices adiabatically with the initial detuning of $-200E_{\rm{r}}$ (with the recoil energy $E_{\rm{r}}\approx 2\pi\times 3.7$kHz).
The parameters $t_0=0.094(2)E_{\rm{r}}$ and $t_{\rm{so}}=0.051(1)E_{\rm{r}}$ are fixed throughout this work~\cite{supMat}.
The huge initial detuning $\delta$  is much greater than the hopping coefficients $t_{0,\rm{so}}$, which suppresses the Raman processes and guarantees that the atoms are prepared in the polarized $\mid\uparrow\rangle$ state~\cite{uncover_topology}.
Secondly, the detuning is quenched from $-200E_{\rm{r}}$ to $\delta\in [-1,1]E_{\rm{r}}$ within 200ns, which activates a non-equilibrium evolution of spin oscillations
between $\left|\uparrow\right\rangle$ and $\left|\downarrow\right\rangle$ states governed by $\mathcal{H}(\bm{q})$.
Finally, after the spin oscillation for a certain time $t$, $P_{x,y,z}^{\rm{exp}}(\bm{q},t)$ are measured by a Bloch state tompgraphy~\cite{supMat}.
To be concrete, the component $P_{z}^{\rm{exp}}(\bm{q},t)$ is directly measured by the spin-resolved time-of-flight (ToF) imaging; after rotating the measurement basis via a momentum-transfer $\pi/2$ Raman pulse, the component $P_{x,y}^{\rm{exp}}(\bm{q},t)$ is also extracted by the spin-resolved ToF imaging.

To determine the CS invariant from the oscillation of $P_{x,y,z}^{\rm{exp}}(\bm{q},t)$,
the following ingredients should be well performed for our apparatus: ultra-low noise of the bias magnetic field $\bm{B}$~\cite{XXT2019}, well-controlled temperature and atomic number of the system and coherent Raman pulse~\cite{yi2023extracting,supMat}.
The root mean square value of the noise of $|\bm{B}|=23.4$ G is strongly suppressed to below $50\mu$G, which enhances the coherence time of evolution of $P_{x,y,z}^{\rm{exp}}(\bm{q},t)$.
For controlling the filling rate of the lowest-band well (depressing the atoms populated in the higher band), the temperature and atomic number are carefully controlled around 100nK and $2\times 10^5$ with stability of 7nK and 6\%, respectively, which further increases the coherence time.
Besides, the relative phase between Raman pulse and Raman lattices is fixed to the specific values with stability of $0.01\pi$, which ensures the acquisition of the spin oscillation~\cite{supMat}.

\begin{figure}[t]
\begin{center}
\includegraphics[width=1\linewidth]{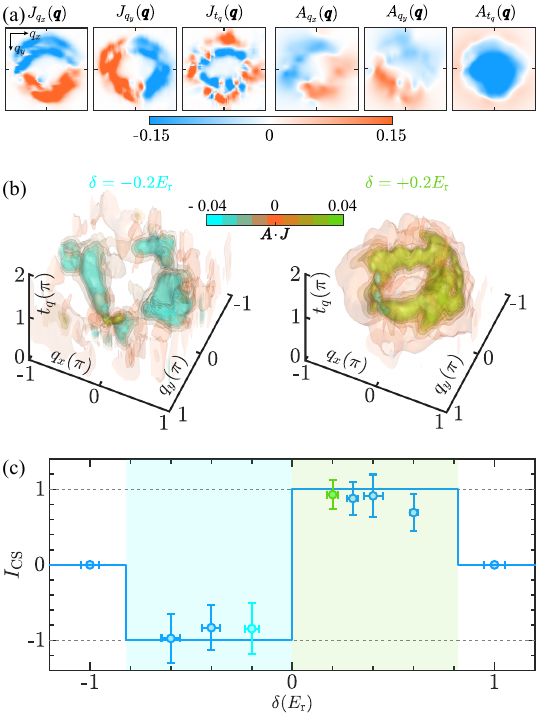}
\caption{Extracting the CS invariant.
(a) The momentum distribution of all the components of $\bm{J}(\bm{k})$ and $\bm{A}(\bm{k})$ at $t_q=0.8\pi$.
(b) The isosurface image of $\bm{A}\cdot\bm{J}$ in the $\bm{k}$ space for $\delta=-0.2E_{\rm{r}}$ (left) and $\delta=+0.2E_{\rm{r}}$ (right).
(c) The CS invariant as a function of the detuning.
The circles with horizontal and vertical error bars (solid lines) are obtained from the experimental measurements and error propagation based on the experimental measurements (the theoretical calculations), respectively.
}
\label{Fig3}
\end{center}
\end{figure}

After tackling the above-mentioned ingredients, 
the target observables $P_{x,y,z}^{\rm{exp}}(\bm{q},t)$ are measured as shown in Fig.~\ref{Fig2}.
In the left of Fig.~\ref{Fig2}, we depict the time-dependent $P_{x,y,z}^{\rm{exp}}(\bm{q})$ that are distributed in the lattice Brillouin zone with more than $10^4$ pixels.
The evolution of $P_{x,y}^{\rm{exp}}(\bm{q})$ forms a spiral-like pattern.
Note that all $P_{x}^{\rm{exp}}(\bm{q})$ roughly coincide with $P_{y}^{\rm{exp}}(\bm{q})$ after $P_{x}^{\rm{exp}}(\bm{q})$ is rotated clockwise by $90^{\circ}$.
In contrast to $P_{x,y}^{\rm{exp}}(\bm{q})$, the evolution of $P_{z}^{\rm{exp}}(\bm{q})$ forms a ring pattern with $C_4$ symmetry.
Further, the time evolution of $P_{x,y,z}^{\rm{exp}}(\bm{q})$ at each quasi-momentum point in the lattice Brillouin zone are extracted.
An example at the quasi-momentum point $(q_x,q_y)=(0.39,0.33)\pi$ is shown in Fig.~\ref{Fig2}.
We observe the oscillation of $P_{x,y,z}^{\rm{exp}}(\bm{q})$, which is fitted by sinusoidal function with damping~\cite{supMat}.
The oscillation frequencies $f$ of the fitting are all around 900Hz, which is consistent with Eq.~(\ref{Eq:Hopf_map_T3_S2}).
However, such data feature a damped oscillation as $t$ increases, which leads to decreasing norms of $P_{x,y,z}^{\rm{exp}}(\bm{q})$ and then deviates from the descriptions of unitary evolution as in Eq.~(\ref{Eq:Hopf_map_T3_S2}).
The damped behaviour may originate from the internal state relaxation from higher bands to lower bands and the noise of the bias magnetic field~\cite{uncover_topology,damp_Rabi1}.
Under the circumstances, we normalize $\bm{P}^{\rm{exp}}(\bm{q},t)$ by $\bm{P}(\bm{q},t)=\bm{P}^{\rm{exp}}(\bm{q},t)/|\bm{P}^{\rm{exp}}(\bm{q},t)|$.
The components of the normalized expectation values of three Pauli matrices $P_{x,y,z}(\bm{q},t)$ all possess the periodicity with stronger amplitudes (See data of green squares with error bars in Fig.~\ref{Fig2}).

Now, we extract quantitatively the CS invariant based on Eq.(\ref{CSinvarint}).
From the measurements of the three components of $P_{x,y,z}(\bm{q},t_q)$ ($t$ has been rescaled to $t_q=2\pi ft$), the components of $\bm{J}(\bm{k})$ are obtained via ${J}_{\mu}(\bm{k})=\epsilon_{\mu \nu \lambda}\bm{P}(\bm{k})\cdot \left [\partial_{\nu}\bm{P}(\bm{k}) \times \partial_{\lambda}\bm{P}(\bm{k})\right ]/(8\pi)$ ~\cite{Wilczek_Zee1983,WU1984325,linking_HuiZhai}, and then the components of $\bm{A}(\bm{k})$ with the gauge $\partial_{\mu}A_{\mu}=0$ are reconstructed from ${J}_{\mu}(\bm{k})$.
All the components of the Berry curvature and Berry connection with typical $t_q$ are exhibited in Fig.~\ref{Fig3}(a).
The distributions of the each component of $\bm{J}(\bm{k})$ and $\bm{A}(\bm{k})$ inherit the similar behaviour as the distributions of $P_{x,y,z}^{\rm{exp}}(\bm{k})$ in Fig.~\ref{Fig2}: the $q_y$-components of both $\bm{J}(\bm{k})$ and $\bm{A}(\bm{k})$ roughly coincide with the corresponding $q_x$-components rotated clockwise by $90^{\circ}$; the $t_q$-components of both $\bm{J}(\bm{k})$ and $\bm{A}(\bm{k})$ possess $C_4$ symmetry.
Such inheritance indicates that, after differentiating and integrating $\bm{P}(\bm{q},t)$, rich distributions of $\bm{J}(\bm{k})$ and $\bm{A}(\bm{k})$ that contain non-trivial properties are still acquired.

We go one step further to directly evaluate the product of $\bm{A}(\bm{k})$ and $\bm{J}(\bm{k})$ according to the experimental measurements, which are displayed in Fig.~\ref{Fig3}(b).
The nonzero values of $\bm{A}\cdot\bm{J}$ are basically distributed in a ring of a finite radius.
The sign of the CS invariant can be determined immediately from
the distribution of $\bm{A}\cdot\bm{J}$: $I_{\rm{CS}}<0~(>0)$ for $\delta=-0.2E_{\rm{r}}~(0.2E_{\rm{r}})$ since most of $\bm{A}\cdot\bm{J}$ take negative (positive) values.
By then summing the distribution of $\bm{A}\cdot\bm{J}$ over the full $\bm{k}$ space, we obtain $I_{\rm{CS}}=0.92\pm0.18$ ($I_{\rm{CS}}=-0.85\pm0.34$) with $\delta=0.2E_{\rm{r}}$ ($\delta=-0.2E_{\rm{r}}$).
In addition, the distribution of $\bm{A}\cdot\bm{J}$ with $|\delta|>8t_0$ is not shown as maximum value of $|\bm{A}\cdot\bm{J}|<10^{-3}$, which indicates $|I_{\rm{CS}}|\rightarrow 0$.
Although local structures of these distributions somewhat deviate from the numerical calculations (See the right of Fig.~\ref{Fig1}) and
the integration over $3\times 10^5$ data points yields values clustering near $\pm1$ or 0, which manifests that the CS invariant is a global quantity and is quite insensitive to the details of the distribution of $\bm{A}\cdot\bm{J}$.

Finally, by altering the detuning $\delta\in [-1,1]E_{\rm{r}}$, the CS invariants exhibits three nearly quantized values of $\pm 1$ and 0, as shown in Fig.~\ref{Fig3}(c).
The experimental measurements of the CS invariants are in agreement with
the theoretically quantized values.
Moreover, the transitions among these values indicate the change of the topology of the quantum state in (2+1)D, which is consistent with our previous measurements of the ground-state Chern number in 2D~\cite{yi2023extracting} and validates the theoretical prediction~\cite{linking_HuiZhai}.
These results also reflect our apparatus possesses powerful capabilities of the control and detection for a topological quantum simulator.

\begin{figure}
\begin{center}
\includegraphics[width=\linewidth]{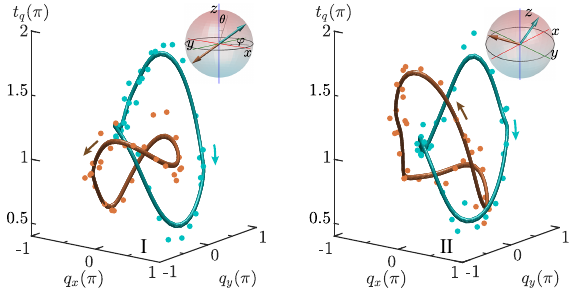}
\caption{Arbitrary pairs of the oriented closed loops with $\delta=0.2E_{\rm{r}}$.
The dots are the experimental data, and the closed curves are obtained by fitting the experimental data.
The location of the closed loops mapping to Bloch sphere are marked by the constant vectors on the Bloch sphere (arrows).
The direction of the closed loops are marked by the arrows near the loops.
The selected loops:
\uppercase\expandafter{\romannumeral1}: $(\theta,\varphi)=(0.3,0.25)\pi$ (cyan) and $(0.7,1.25)\pi$ (brown);
\uppercase\expandafter{\romannumeral2}: $(\theta,\varphi)=(0.3,0)\pi$ (cyan) and $(0.4,0.75)\pi$ (brown).
}
\label{Fig4}
\end{center}
\end{figure}

In addition, the CS invariant can be interpreted by a non-trivial linking structure, which is equal to the linking number between oriented closed loops~\cite{PhysRevA.64.052101,linking_HuiZhai,link2019}.
The linking number is nonzero (vanishes) when a pair of the oriented closed loops in $T^3$ is (is not) interlinked~\cite{linking_HuiZhai}.
An oriented closed loop in $T^3$ corresponds to a constant vector on the Bloch sphere $S^2$ (Fig.\ref{Fig4}), which is given by the map from $\bm{P}(\bm{q},t_q)$ in $T^3$ to the constant vectors on $S^2$, i.e., $\bm{P}(\bm{q},t_q)=(\sin\theta\cos\varphi, \sin\theta \sin\varphi, \cos\theta)$~\cite{Hopf1931,Moore-Wen2008,linking_HuiZhai}.
Here, $\theta$ and $\varphi$ are respectively the polar angle and the azimuthal angle defined in $S^2$, as shown in Fig.~\ref{Fig4}.
Hereby, arbitrary pairs of the closed loops that satisfy such map can be extracted.
Two typical pairs of the closed loops are plotted in Fig.~\ref{Fig4}.
No matter what values of the angle between two constant vectors on $S^2$ are taken, each pair of the closed loops is interlinked and thus the number of the linking is equal to unity.
According to the right-band rule \cite{supMat}, we obtain the linking number is +1 for arbitrary pairs of closed loops, which implies $I_{\rm{CS}}=+1$, in consistent with the integral of $\bm{A}\cdot \bm{J}$.

In summary, the CS invariant is measured on account of the measurements of the Berry curvature and the corresponding Berry connection.
Furthermore, the present system can be extended to study 3D topological models with non-Abelian CS invariants, which apply to systems with degenerate energy bands and are responsible to exotic transport phenomena termed topological magneto-electric effect~\cite{PhysRevB.78.195424,Chiu2016}.
In Ref. \cite{supMat}, we propose two experimental schemes.
First, we can map the quench processing of the present two-band model to four-band model, and then measure transport coefficient of the four-band model, which is given by non-Abelian CS invariant.
Second, based on 3D Raman lattices that we have implemented~\cite{wang2021realization}, we plan to construct a genuine 3D four-band time-reversal-invariant topological insulator featuring nontrivial 3D non-Abelian CS invariant.

\begin{acknowledgments}
We thank Hui Zhai for fruitful discussions.
This work was supported by the Innovation Program for Quantum Science
and Technology (Grant No.2021ZD0302001 and 2021ZD0302100), the National Natural Science Foundation of China (Grant No.12025406, 12374248, 12104445 and 12374478), Anhui Initiative in Quantum Information Technologies (Grant No.AHY120000), Shanghai Municipal Science and Technology Major Project (Grant No.2019SHZDZX01), and the Strategic Priority Research Program of Chinese Academy of Science (Grant No.XDB28000000). J.Z. acknowledges support from the CAS Talent Introduction Program (Category B) (Grant No.KJ9990007012) and the Fundamental Research Funds for the Central Universities (Grant No.WK9990000122). J.Y. acknowledges support from the Innovational Fund for Scientific and Technological Personnel of Hainan Province (Grant No. KJRC2023B11).
\end{acknowledgments}


\newpage
\onecolumngrid
\renewcommand\thefigure{S\arabic{figure}}
\setcounter{figure}{0}
\renewcommand\theequation{S\arabic{equation}}
\setcounter{equation}{0}
\makeatletter
\newcommand{\rmnum}[1]{\romannumeral #1}
\newcommand{\Rmnum}[1]{\expandafter\@slowromancap\romannumeral #1@}
\makeatother

\newpage

{
\center \bf \large
Supplemental Material for: \\
Supplemental Materials for: Measuring the Chern-Simons invariant in quantum gases\vspace*{0.1cm}\\
\vspace*{0.0cm}
}

\vspace{4ex}

\maketitle

In this supplemental materials, we describe the experimental setup, Bloch-state tomography in a quenched Raman lattice, the experimental protocol, the fitting of experimental data, the determination of the sign of the linking number, improvement of experimental technology, the relationship between CS invariant and Chern number and experimental proposals for realizing topological models with non-Abelian CS invariants using quantum gases.

\subsection{The experimental setup}\label{lab1}

\begin{figure}
\begin{center}
\includegraphics[width=0.6\linewidth]{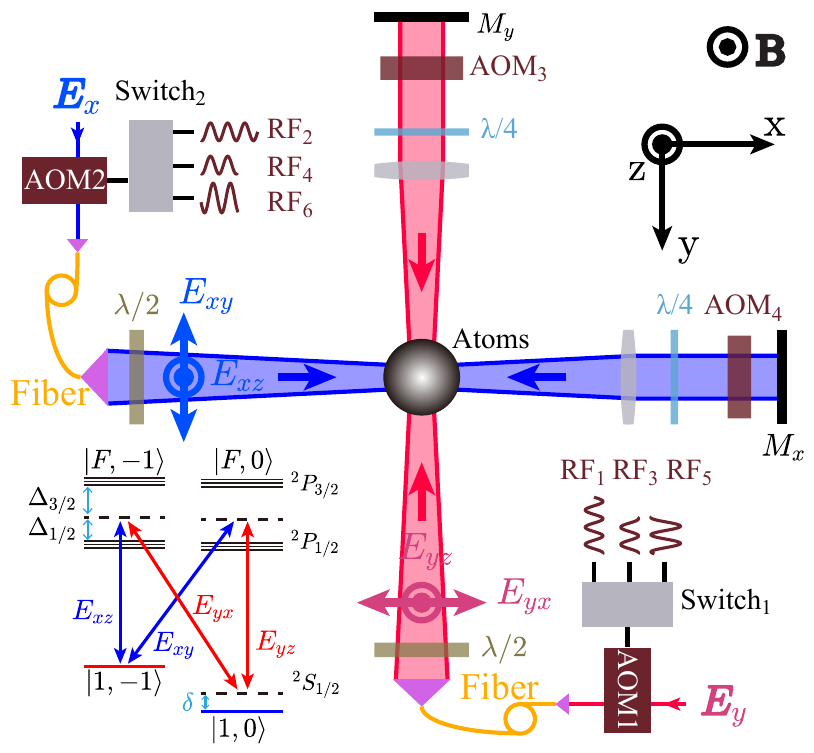}
\caption{The experimental setup.
The two laser beams $\bm{E}_{x,y}$, being splitted into two orthogonality polarized components $\bm{E}_{x,y}=\bm{E}_{xy,yx}+\bm{E}_{xz,yz}$ by $\lambda/2$ waveplates, are used to generate lattice potentials, Raman potentials and Raman pulse.
$\lambda/4$ waveplates are used to generate topological non-trivial Hamiltonian.
$M_{x,y}$ are mirrors.
$\text{AOM}_{3,4}$ are used to tune the the strength of the beams $\bm{E}_{x,y}$ reflected back to atomic cloud.
$\text{AOM}_{1,2}$ control the frequency, intensity and phase of the beams $\bm{E}_{x,y}$ via ratio frequency signals $\text{RF}_{1,2,3,4,5,6}$.
The bias magnetic field $\bm{B}=B\hat{z}$ along $\hat{z}$ direction.
Inset: level structures and two Raman couplings $\Omega_{1,2}\propto |\bm{E}_{xy,yx}||\bm{E}_{yz,xz}|$.
}
\label{quenchTomographyScheme}
\end{center}
\end{figure}

The experimental setup is shown in Fig.\ref{quenchTomographyScheme}.
The laser beam $\bm{E}_x$ ($\bm{E}_y$) with the wavelength $\lambda=$787nm along $\hat{x}$ ($\hat{y}$) direction, being splitted into two orthogonality polarized components $\bm{E}_{x}=\bm{E}_{xy}+\bm{E}_{xz}$ ($\bm{E}_{y}=\bm{E}_{yx}+\bm{E}_{yz}$) by a $\lambda/2$ waveplate, is irradiated into the ultracold atomic cloud of ${}^{87}$Rb.
Then, the laser beam is reflected back to the atomic cloud by the mirror $M_{x}$ ($M_y$).
Meanwhile, the beam passes through the $\lambda/4$ waveplate and acousto-optic modulator $\text{AOM}_3$ ($\text{AOM}_4$).
The $\lambda/4$ waveplate induces a phase shift of $\pi$ between $\bm{E}_{xy}$ ($\bm{E}_{yx}$) and $\bm{E}_{xz}$ ($\bm{E}_{yz}$), which plays a key role in generating the topological Hamiltonian of Raman lattices.
The acousto-optic modulators $\text{AOM}_{3,4}$ are used to adjust the ratio of the beam reflected back to the atomic cloud, marked as $\gamma$.
To maintain the polarization of the laser through the $\text{AOM}_{3,4}$, we fabricate $\text{AOM}_{3,4}$ with an acousto-optical material called ``dense flint glass''.
Furthermore, two magnetic sublevels $|F=1,m_F=-1\rangle$ (defined as spin up $|\uparrow\rangle$) and $|F=1,m_F=0\rangle$ (defined as spin down $|\downarrow\rangle$) have a Zeeman splitting of 16.5MHz induced by a bias field $\bm{B}=B\hat{z}$ with the strength $B=23.4$G, and are coupled by two Raman transitions $\Omega_{1,2}$.
Thus, the Hamiltonian with the pseudo-spin 1/2 is written as
\begin{equation}\label{Ramanlattice}
H=\begin{pmatrix}
\bm{p}^2/2m+\gamma V_{\rm{latt}}(x,y)+\delta/2 & \Omega_{\rm{R}}(\gamma)\\
\Omega_{\rm{R}}^*(\gamma) & \bm{p}^2/2m+\gamma V_{\rm{latt}}(x,y)-\delta/2
\end{pmatrix},
\end{equation}
where $\bm{p}=(p_x,p_y)$, $m$ and $\delta$ is respectively the momentum, the atomic mass and two-photon detuning.
The Raman potential $\Omega_{\rm{R}}(\gamma)$ is a function of $\gamma$ (See Ref. \cite{yi2023extracting} for details).
Regarding $\gamma=1$, the system of the 2D Raman lattices is hold.
The optical lattices $V_{\rm{latt}}(x,y)=V_0[\cos^2(k_0x)+\cos^2(k_0y)]$ with lattice depth $V_0$.
The Raman potentials $\Omega_{\rm{R}}=\Omega_1(x,y)-i\Omega_2(x,y)$ with $\Omega_{1}(x,y)=\Omega_0\cos(k_0x)\sin(k_0y)$ and $\Omega_2(x,y)=\Omega_0\cos(k_0y)\sin(k_0x)$.
Here, $\Omega_0$ is the Raman coupling strength.
Under the two-band tight-binding approximation, the Hamiltonian~(\ref{Ramanlattice}) has the form of QAH model, i.e. $\mathcal{H}(\bm{q})=\bm{h}\cdot \bm{\sigma}=2t_{\rm{so}}\sin(q_y)\sigma_x+2t_{\rm{so}}\sin(q_x)\sigma_y+[\delta/2-2t_0(\cos(q_x)+\cos(q_y))]\sigma_z$, which is same as the main text.
And the hopping coefficients $t_0$ and $t_{\rm{so}}$ are respectively determined by the lattice depth $V_0$ and the Raman coupling strength $\Omega_0$~\cite{realization2DSOC,uncover_topology}.
%
The typical experimental parameters $V_0=4E_{\rm{r}}$ and $\Omega_0=1E_{\rm{r}}$ correspond to $t_0\approx0.09E_{\rm{r}}$ and $t_{\rm{so}}\approx0.05E_{\rm{r}}$, respectively.

Regarding $\gamma=0$, the Raman pulse, being used to rotate the measurement basis and transfer the momentum during Bloch state tomography, can be realized.
Then, the lattice potential $V_{\rm{latt}}=0$.
The Raman potentials turn into the Raman coupling $\Omega_{\rm{R}}=\Omega_{\text{R0}}e^{-i[k_0(x+y)-\Delta\varphi]}$.
Here, $\Omega_{\text{R0}}=\sqrt{2}\Omega_0/4$ is Raman pulse strength and $\Delta\varphi$ is the relative phase between Raman pulse and Raman lattices~\cite{yi2023extracting}.
In order to distinguish from the Hamiltonian of the Raman lattices, we express the Hamiltonian of the Raman pulse as
\begin{equation}\label{Ramanpulse}
H_{\rm{R}}=\begin{pmatrix}
  \bm{p}^2/2m+\delta_{\rm{R}}/2& \Omega_{\rm{R}}\\
  \Omega_{\rm{R}}^*& \bm{p}^2/2m-\delta_{\rm{R}}/2
\end{pmatrix},
\end{equation}
where $\delta_{\rm{R}}$ is the two-photon detuning of the Raman pulse.

\subsection{Bloch state tomography in a quenched Raman lattice}~\label{lab2}
The main idea of the Bloch state tomography is to rotate the momentum dependent measurement basis to obtain the expectation value of three Pauli matrices $P_{x,y,z}(\bm{q},t)$ using a Raman pulse, as shown in Fig.~\ref{timeSerialAndPhysicalTomography}(b).
$P_{z}(\bm{q},t)$ is obtained by a direct spin-resolved time-of-flight imaging;
after rotating the measurement basises $|\bm{q},\uparrow\rangle$ and $|\bm{q}+\bm{k}_0,\downarrow\rangle$ to $x$ ($y$) axis, $P_{x}(\bm{q},t)$ ($P_{y}(\bm{q},t)$) is extracted by the spin-resolved time-of-flight imaging.

In order to obtain the expression to extract $P_{x,y,z}(\bm{q},t)$, we consider the evolution of the state and fix the measurement basis during the quench process and the Raman pulse.
After preparing the initial state $\mid\uparrow\rangle=(1~0)^T$ ($T$ stands for transpose.) in the deep topologically trivial regime with $\delta\gg t_{0,\rm{so}}$, we quench the system to a the regime with $|\delta|\sim t_{0,\rm{so}}$, thus the initial state evolve to a superposition state $|\Psi(\bm{q},t)\rangle=\exp(-iHt)\mid\uparrow\rangle$ after holding a certain time $t$.
Subsequently, a Raman pulse with pulse time $t_{\rm{R}}$ is applied to rotate the superposition state and the rotated state is expressed as $| \tilde{\Psi}(\bm{q},t,t_{\rm{R}})\rangle=\exp(-iH_{\rm{R}}t_{\rm{R}}) |\Psi(\bm{q},t)\rangle$.
Finally, using spin-resolved time-of-flight imaging, the time-dependent expectation value of three Pauli matrices are extracted from the equation (see Ref.~\cite{yi2023extracting})
\begin{equation}\label{Eqtomography}
\begin{aligned}
P(\bm{q},t)=&\langle \tilde{\Psi}(\bm{q},t,t_{\rm{R}}) \mid \sigma_z \mid \tilde{\Psi}(\bm{q},t,t_{\rm{R}}) \rangle=P_z(\bm{q},t) \cos(2\Omega_{\rm{R0}}t_{\rm{R}})\\
&+(P_x(\bm{q},t) \sin\Delta\varphi+P_y(\bm{q},t) \cos\Delta\varphi) \sin(2\Omega_{\rm{R0}}t_{\rm{R}})
\end{aligned}
\end{equation}
after neglecting the kinetic energy term $\bm{p}^2/2m$ and setting $\delta_{\rm{R}}=0$ in the Hamiltonian $H_{\rm{R}}$.
For $t_{\rm{R}}=0$, $P_z(\bm{q},t)$ is obtained;
for $t_{\rm{R}}=\pi/(4\Omega_{\rm{R0}})$ (i.e. a $\pi/2$ Raman pulse) and $\Delta\varphi=\pi/2~(0)$, $P_x(\bm{q},t)$ ($P_y(\bm{q},t)$) is extracted.
For two band QAH model, the time-dependent expectation value of three Pauli matrices satisfies $\sqrt{P_x^2(\bm{q},t)+P_y^2(\bm{q},t)+P_z^2(\bm{q},t)}=1$.

\subsection{The experimental protocol}\label{lab3}

\begin{figure}
\begin{center}
\includegraphics[width=0.6\linewidth]{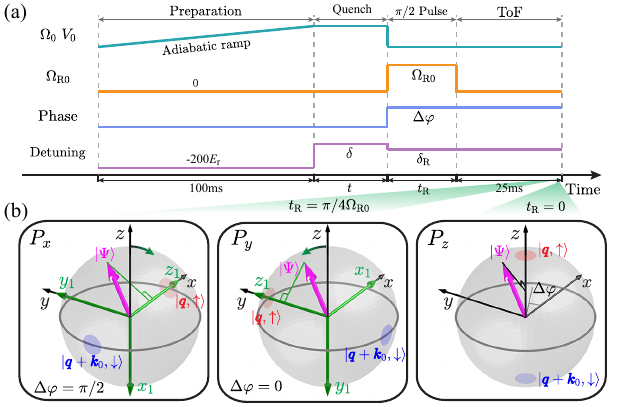}
\caption{The experimental protocol and the rotation of the measurement basis.
(a) The experimental protocol.
The atoms is firstly prepared in a spin polarized state by ramping optical lattice depth to $V_0=4.0E_{\rm{r}}$ and Raman strength to $\Omega_0=1.0E_{\rm{r}}$, and setting the detuning as $\delta=-200E_{\rm{r}}$.
Subsequently, the detuning is suddenly switched close to resonance, e.g., $\delta=0.2E_{\rm{r}}$.
After holding a certain time $t$, the Raman pulse with pulse duration $t_{\rm{R}}$ is switched on.
When $t_{\rm{R}}=0$, $P_z(\bm{q},t)$ is measured directly after spin-resolved TOF imaging;
When $t_{\rm{R}}=\pi/(4\Omega_{\rm{R0}})$ and $\Delta\varphi=\pi/2 (0)$, $P_x(\bm{q},t)$ ($P_y(\bm{q},t)$) is obtained after spin-resolved TOF imaging.
(b) The rotation of the measurement basis after the Raman pulse.
In order to obtain $P_x(\bm{q},t)$ ($P_y(\bm{q},t)$), the measurement basises $|\bm{q},\uparrow\rangle$ and $|\bm{q}+\bm{k}_0,\downarrow\rangle$ are rotate to $x$ ($y$) axis (black arrows), and the rotated axes are redefined as $x_1,y_1,z_1$ (green arrows).
The projection to the $z_1$ ($z$) axis is realized by spin-resolved TOF imaging.
The magenta arrows indicate the state $|\Psi(\bm{q},t)\rangle$ to be measured.
}
\label{timeSerialAndPhysicalTomography}
\end{center}
\end{figure}

The all components of the time-dependent expectation value of the three Pauli matrices $P_{x,y,z}(\bm{q},t)$ is measured by Bloch state tomography in a quenched two-dimensional Raman lattice.
The experimental protocol is performed as follows (see Fig.~\ref{quenchTomographyScheme} and Fig.~\ref{timeSerialAndPhysicalTomography}):

1) \emph{The preparation of the initial state}.
The atoms are adiabatically loaded into the Raman lattices in 100ms with the temperature of 100nK, which causes atoms to populate each momentum point in the first Brillouin zone.
In the meanwhile, the depth of the optical lattice potential and the strength of the Raman potentials is respectively ramped from 0 to $V_0=4.0E_{\rm{r}}$ and $\Omega_0=1.0E_{\rm{r}}$ by increasing the intensity of the beams $\bm{E}_{x,y}$.
The detuning is set as $\delta=-200E_{\rm{r}}$ by controlling the frequency of the beams.
The intensity and frequency of the beams are tuned by opening ratio-frequency (RF) signal $\rm{RF}_{1,2}$.
In the meanwhile, other RF signals and $\rm{AOM}_{3,4}$ are closed.
Note that $\delta \gg \Omega_0$ suppresses the Raman potentials and spin flipping vanishes, thus the atoms are in the spin polarized state $\mid \uparrow \rangle$.

2) \emph{The quenching and the evolution of the state}.
The quench is executed by switching the detuning from $-200E_{\rm{r}}$ to $\delta\in [-1,1]E_{\rm{r}}$ within 200ns.
The sudden quench, being much less than the characteristic time of the state evolution (about several hundred microseconds), is controlled by RF switches.
Meanwhile, $\rm{RF}_{3,4}$ control the intensity and the frequency of the beams during the quench.
Note that the initial phase of $\rm{RF}_{3,4}$ must be fixed so that the relative phase $\Delta\varphi$ between the Raman pulse and the Raman lattices maintains a definite value.
Other RF signals are turned off in this step.
After the quench, the state evolves out of equilibrium and oscillates between $\mid \uparrow \rangle$ and $\mid \downarrow \rangle$.

3) \emph{The detection of the state using Bloch state tomography}.
After the quench for a certain time $t$, a $\pi/2$ Raman pulse with duration $t_{\rm{R}}$ is applied to the atoms within 200ns via the following manipulation~\cite{yi2023extracting}: (i) Turning off the retroreflective beams of $\bm{E}_{x,y}$ via opening $\rm{AOM}_{3,4}$;
(ii) Turning on the strength of Raman pulse to $\Omega_{\text{R0}}\approx 3.4E_{\rm{r}}$ by tuning the intensity of $\bm{E}_{x,y}$;
(iii) Setting the relative phase $\Delta\varphi$ to a certain value by tuning the phase of $\bm{E}_{x}$ or $\bm{E}_{y}$;
(iv) Setting $\delta_{\text{R}}=0$ by adjusting the frequency of $\bm{E}_{x}$ or $\bm{E}_{y}$.
Meanwhile, the intensity, frequency and phase of the beams are tuned by $\rm{RF}_{5,6}$.
$\rm{RF}_{1,2}$ are turned off and $\rm{RF}_{3,4,5,6}$ are turned on.
Subsequently, the spin-resolved TOF imaging is used to obtain the spin texture $P^{\rm{exp}}(\bm{q},t)=[n_{\uparrow}(\bm{q},t)-n_{\downarrow}(\bm{q},t)]/[n_{\uparrow}(\bm{q},t)+n_{\downarrow}(\bm{q},t)]$, where $n_{\uparrow}$ ($n_{\downarrow}$) represents the time-dependent atomic density of $\mid \uparrow \rangle$ ($\mid \downarrow \rangle$) in the momentum space.
$P_z^{\rm{exp}}(\bm{q},t)$ is obtained when $t_{\rm{R}}=0$, as has been applied in Ref.~\cite{realization2DSOC,uncover_topology}.
And $P_x^{\rm{exp}}(\bm{q},t)$ ($P_y^{\rm{exp}}(\bm{q},t)$) is detected when $t_{\rm{R}}=10\mu s$ and $\Delta\varphi=\pi/2~(0)$.
The pulse duration $t_{\rm{R}}$ is short enough so that the atoms hardly evolve during this Raman pulse.
The relative phase $\Delta\varphi$ is tuned by the phase of $\rm{RF}_5$ or $\rm{RF}_6$.

\subsection{Fitting the experimental data}\label{lab4}
We firstly fit the experimental data of $P_{x,y,z}^{\rm{exp}}(\bm{q},t)$ with a general function~\cite{damp_Rabi1}
\begin{equation}\label{fitFun}
F(\bm{q},t)=Ae^{-t/\tau_1}\cos(2\pi ft+\phi)+Be^{-t/\tau_2}+C,
\end{equation}
which has been used in Ref.~\cite{uncover_topology,Yi2019}.
Here, the first term is the damped oscillation induced by the dephasing with the amplitude $A$, the characteristic dephasing time $\tau_1$, the oscillation frequency $f$ and the phase $\phi$, the second term is the internal relaxation with the amplitude $B$ and the characteristic decay time $\tau_2$, and the third term is the offset $C$.
An example of the fitting is shown in Fig. 2 of the main text.
Therefore, we normalize $P_{x,y,z}^{\rm{exp}}(\bm{q},t_q)$ by $P_{x,y,z}(\bm{q},t_q)=P_{x,y,z}^{\rm{exp}}(\bm{q},t_q)/|\bm{P}^{\rm{exp}}(\bm{q},t_q)|$, which is used to calculate the CS invariant (See Fig.~3) and extract all the closed loops according to the map $\bm{P}(\bm{q},t_q)=(\sin\theta\cos\varphi, \sin\theta \sin\varphi, \cos\theta)$ (See Fig.~4).
An example for extracting a point on a closed loop is shown in Fig.~2 of the main text.
In addition, the fitting is achieved by the least square method and the errors of the spin oscillation at each time measured by the experiment are introduced into the weight of the fitting.
Using the fitting, a jacobian matrix J and chi-square C can be obtained.
Eventually, the errors of all fitting parameters are given by $\sqrt{C(JJ^T)^{-1}}$.

\subsection{Determining the sign of the linking number}\label{lab5}
\begin{figure}
\begin{center}
\includegraphics[width=0.5\linewidth]{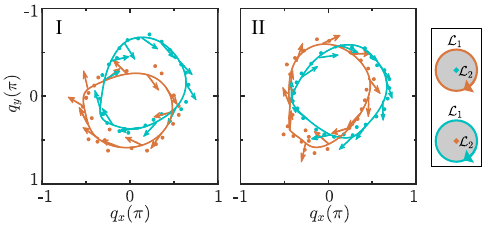}
\caption{The direction of the oriented closed loops in Fig. 4.
The vertical view of two pairs of the closed loops.
The arrows denote the tangent directions at the quasi-momentum points on the loops determined by the Berry curvature vector.
The length of the arrows have been normalized.
The dots are the experimental data, and the curves are obtained by fitting the experimental data.
The inset is a schematic representation of a pair of the loops $\mathcal{L}_{1,2}$, the dots indicate that $\mathcal{L}_2$ points out of the paper.
The gray regions are the surfaces enclosed by $\mathcal{L}_1$.
}
\label{FigFiberDirection}
\end{center}
\end{figure}

We determine the sign of the linking number by the right-hand rule according to the orientation of any pair of the closed loops.
If the orientation of any pair of the closed loops follows the right-hand rule, the sign of the linking number is negative; otherwise, the sign is positive.
The orientation of each loop is determined by the Berry curvature $\bm{J}_{\mu}(\bm{k})=\epsilon_{\mu \nu \lambda}\bm{P}(\bm{k})\cdot \left[\partial_{\nu}\bm{P}(\bm{k}) \times \partial_{\lambda}\bm{P}(\bm{k})\right]$ \cite{linking_HuiZhai,Hopflinks}.
Figure.~\ref{FigFiberDirection} shows the Berry curvature $\bm{J}(\bm{k})$ on the tangent directions of the closed loops in ($q_x-q_y$) plane, indicating that the direction of each loop is counterclockwise.
Besides, we find that the brown (cyan) loop (being named as $\mathcal{L}_2$) always crosses the surface enclosed by the cyan (brown) loop (being named as $\mathcal{L}_2$), and the direction of $\mathcal{L}_2$ at the intersection between $\mathcal{L}_2$ and the surface enclosed by $\mathcal{L}_1$ points out of the paper, the geometric relationship of which is abstracted to the far right of Fig.~\ref{FigFiberDirection}.
Now, we apply the right-hand rule according to the far right of Fig.~\ref{FigFiberDirection}.
One aligns the bending orientation of the four fingers with the orientation of $\mathcal{L}_1$, and the orientation of $\mathcal{L}_2$ is opposite to the orientation of the thumb, indicating the sign of the linking number is positive.
Applying the right-hand rule to any pair of the closed loops, the sign of the linking number of any pair of the loops is positive, indicating the sign of the CS invariant is positive for $\delta=0.2E_{\rm{r}}$.


\subsection{Improvement of experimental technology}\label{lab6}
We need to improve the control of experimental parameters on top of our previous works~\cite{realization2DSOC,W.S_longlive,yi2023extracting} to probe the Chern-Simons invariant.
These parameters include atom number, atom temperature and the relative phase $\Delta \varphi$.

In terms of atom number and atom temperature, these two parameters are set close to $2\times 10^5$ and 100nK, respectively, which are at a similar level with our previous works.
However, to increase the amplitude of spin oscillation, we have optimized the stability of the atom number and atom temperature, which are around 6\% and 7nK, respectively.

\begin{figure}
\begin{center}
\includegraphics[width=1\linewidth]{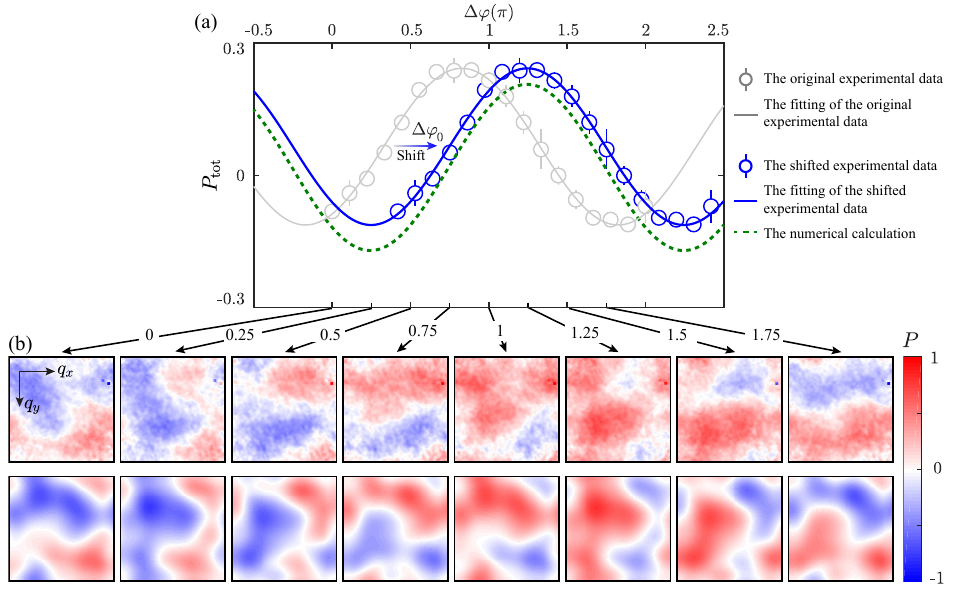}
\caption{The calibration of the relative phase $\Delta\varphi$ between the Raman pulse and the Raman lattices.
(a) The spin polarization $P_{\rm{tot}}$ as a function of $\Delta\varphi$ with $t=150\mu s$.
(b) The spin textures in the $(q_x,q_y)$ space versus the relative phase $\Delta\varphi$.
The first (second) row is experimental measurements (numerical calculations).
}
\label{FigcalibrationPhase}
\end{center}
\end{figure}

A vital improvement in our current work compared with our previous works is the accurate control of the relative phase $\Delta \varphi$.
In the experiment, we find $\Delta \varphi$ changes from 0 to $2\pi$ as the holding time $t$ goes from 0 to 1.5ms (due to the fact that different lengths of RF signals input to the AOM do not produce the same phase delay to the laser beam), which leads to the inability to obtain the signal of the spin oscillation. 
Thus, the relative phase $\Delta\varphi$ must to calibrated and fixed.
To this end, we use the same method as the above experimental protocol (in Sec.\ref{lab3}), {\color{blue}except} that the spin polarization is defined as $P_{\rm{tot}}(\Delta\varphi)=(N_{\uparrow}-N_{\downarrow})/(N_{\uparrow}+N_{\downarrow})$, where $N_{\uparrow}$ ($N_{\downarrow}$) is the total number of the atoms in $\mid\uparrow\rangle$ ($\mid\downarrow\rangle$).
By adjusting the phase of $\text{RF}_6$ from 0 to $2\pi$, we obtain $P_{\rm{tot}}$ [the gray dots in Fig.~\ref{FigcalibrationPhase}(a)], which is fitted using a sinusoidal function with an independent variable $\Delta\varphi$ [the gray curve in Fig.~\ref{FigcalibrationPhase}(a)].
Then, the spin polarization $P_{\rm{tot}}$ as a function of $\Delta\varphi$ are calculated numerically [the green curve in Fig.~\ref{FigcalibrationPhase}(a)].
There exists a phase difference $\Delta\varphi_0$ between the numerical calculations (the green curve) and the experimental measurements (the gray dots).
We shift the experimental measurements $\Delta\varphi_0$ so that the translated experimental measurements (the blue dots and curve) basically coincide with the numerical calculations, from which we can calibrate the relative phase $\Delta\varphi$.
Subsequently, we extract the spin textures at different relative phases, as shown in Fig.~\ref{FigcalibrationPhase}(b).
The experimental measurements are consistent with the numerical calculations.
Therefore, the expectation value of the Pauli matrix $P_{x}(\bm{q},t)$ ($P_{y}(\bm{q},t)$) is determined by the spin texture with $\Delta\varphi=\pi/2$ ($\Delta\varphi=0$).
And the stability of the relative phase $\Delta\varphi$ is $0.01\pi$.
Only after these optimizations do we obtain sufficiently accurate CS invariants.

\subsection{The relationship between CS invariant and Chern number}\label{lab7}

These two topological invariants should be exactly equal to each other because of the theorem proved in Ref.~\cite{linking_HuiZhai}. To be concrete, if one starts with the trivial state of a 2D system and evolved under a post-quench Hamiltonian with Chern number $C$, then one could define a 3D topological invariant $I_{\rm{H}}$ describing the topology of the dynamic quantum state, and one always has $C = I_{\rm{H}}$.
We now know that $I_{\rm{H}}$ in this context is just the simplest abelian form of the 3D CS invariant.
This theorem in Ref.~\cite{linking_HuiZhai} makes a close connect between the ground-state topology of a 2D system and the dynamical-state topology of a (2+1)D system. Our experimental results in Fig. 3(c) together with the ones in Fig. 4(b) of Ref.~\cite{yi2023extracting} directly validates the theoretical prediction of Ref.~\cite{linking_HuiZhai}.

On the other hand, although these two quantities have such a relation in the current context, we should emphasis that they are in fact different topological invariants in different systems within different dimensions. In particular, for the dynamical quantum state activated via parameter quench, we could also define its Chern number at any particular time. However, as the initial state is prepared as the ground state of a trivial polarized state with zero Chern number and unitary evolution will never change this number, the Chern number for the dynamical state at any particular time remains zero. This shows that Chern number and CS invariant is generally different.

What's more, CS invariant in the present system has its unique physical consequence--the linking structure of closed loops in (2+1)D space-time as shown in Fig. 4, which is not shared by any Chern insulators with nontrivial Chern numbers.

%

\subsection{Experimental proposals for realizing topological models with non-Abelian CS invariants using quantum gases}\label{lab8}
The Berry connection and the CS invariant probed in the main texts apply to the case that the energy bands are non-degenerate. 
This Berry connection has a natural non-Abelian generalisation for the case that the energy bands are degenerate, which leads to an non-Abelian CS invariant.
Such non-Abelian CS invariant serves as the strong invariant that classifies the 3D symmetry-protected topological insulators~\cite{Chiu2016}, which leads to the topological magneto-electric effect (TME) as a low-energy simulation of axion electrodynamics~\cite{Wilczek-axion-electrodynamics-1987,PhysRevB.78.195424}.
In this section, we propose two experimental schemes for realizing topological models with non-Abelian CS invariants ~\cite{PhysRevB.78.195424,Chiu2016} using quantum gases: 
(1) As a proposal that can be implemented immediately using the setup of the current work, we map the present quench process of two-band model to four-band topological model with TME, and measure the transport coefficient of the four-band model given by non-Abelian CS invariant.
(2) Further,  based on three-dimensional (3D) Raman lattices that we have implemented~\cite{wang2021realization}, we plan to construct a genuine 3D four-band time-reversal-invariant topological insulator  featuring nontrivial 3D non-Abelian CS invariant.

\subsubsection{Measuring non-Abelian CS invariant based on mapping from the quench process of two-band models to four-band models}
In the main text, our experimental setup focuses on measurement of Abelian CS invariant, but the setup can be generalized in measuring of non-Abelian CS invariant.
Specifically, the quench process of two-band model can be mapped to a four-band topological model with TME, and therefore the transport coefficient of the 3D four-band model is also related to the linking of the quenched spin system. In the topological insulators of 3D, the transport coefficient is given by the non-Abelian CS invariant.

We consider a typical 3D topological model with chiral symmetry within class AIII~\cite{Chiu2016},
\begin{align}
        H = \cos(q_z)\sigma_x\otimes \mathbb{I} -\sin(q_z)\sigma_y \otimes(\bm{\sigma}\cdot \bm{h}(\bm{q}))=\left[\begin{array}{cc}
                0&\exp(iq_z\bm{h}(\bm{q})\cdot\bm{\sigma})\\
                \exp(-iq_z\bm{h}(\bm{q})\cdot\bm{\sigma})&0
        \end{array}\right],
\end{align}
with chiral operator $\sigma_z\otimes\mathbb{I}$. The eigenstates of the Hamiltonian are,
\begin{align}
        &E=1,&&\frac{1}{\sqrt{2}}\left[\begin{array}{c}
|\uparrow\rangle\\
        \exp(-iq_z\bm{h}(\bm{q})\cdot\bm{\sigma})|\uparrow\rangle
        \end{array}\right], &&\frac{1}{\sqrt{2}}\left[\begin{array}{c}
|\downarrow\rangle\\
        \exp(-iq_z\bm{h}(\bm{q})\cdot\bm{\sigma})|\downarrow\rangle
\end{array}\right],\\
        &E=-1,&&\frac{1}{\sqrt{2}}\left[\begin{array}{c}
|\uparrow\rangle\\
        -\exp(-iq_z\bm{h}(\bm{q})\cdot\bm{\sigma})|\uparrow\rangle
        \end{array}\right],&&\frac{1}{\sqrt{2}}\left[\begin{array}{c}
|\downarrow\rangle\\
        -\exp(-iq_z\bm{h}(\bm{q})\cdot\bm{\sigma})|\downarrow\rangle
\end{array}\right].
\end{align}
As can be inferred from the eigenstates above, the final state could be mapped with the quench process. In specific, the analog between quench simulations and four-band chiral topological models is explained as following. $q_x$, $q_y$ and $q_z$ in topological insulators are mapped to $q_x$, $q_y$ and $t$, respectively. Then the mapping from quenching state to eigen state of four-band model is constructed as follows:
\begin{align}
        f(q_x, q_y, q_z):\exp(-iq_z\bm{h}(q_x,q_y)\cdot\bm{\sigma})|\uparrow\rangle\rightarrow \frac{1}{\sqrt{2}}\left[\begin{array}{c}
|\uparrow\rangle\\
                -\exp(-iq_z\bm{h}(q_x,q_y)\cdot\bm{\sigma})|\uparrow\rangle
        \end{array}\right].
\end{align}
The Berry connection is evaluated in the four-band AIII topological model,
\begin{align}
        A_{ss,\mu}(q_x,q_y,q_z)=\frac{i}{2\pi}\langle s|\exp(iq_z\bm{h}(q_x,q_y)\cdot\bm{\sigma})\partial_\mu\exp(-iq_z\bm{h}(q_x,q_y)\cdot\bm{\sigma}) |s\rangle, \;\; ss=\uparrow\uparrow,\uparrow\downarrow,\downarrow\uparrow,\downarrow\downarrow,\label{eq:AdefBerryConn}
\end{align}
which is the same as quench process.

\begin{figure*}[ht]
\includegraphics[width=0.5\textwidth]{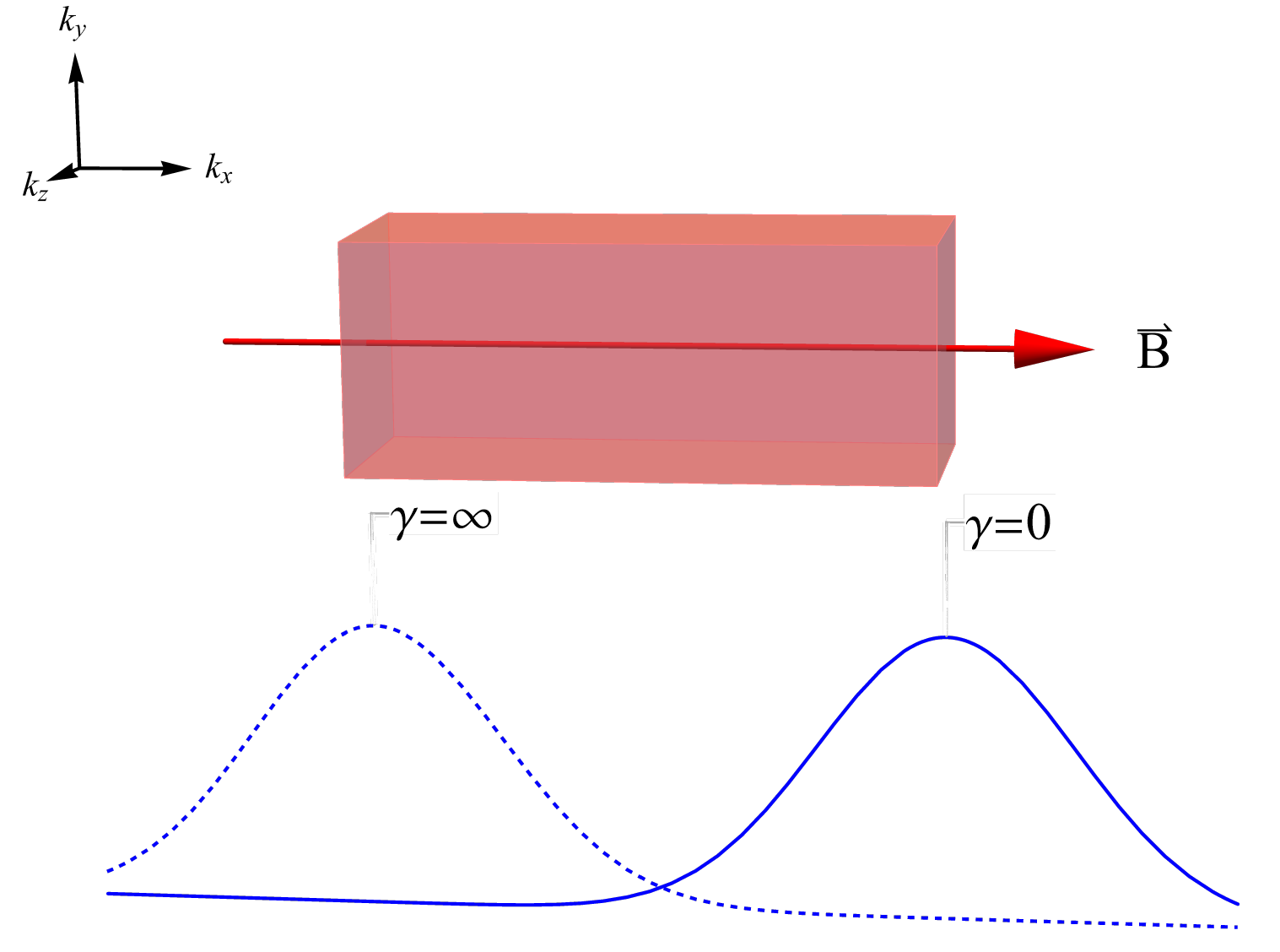}
        \caption{\label{fig:CSchargePump} Illustration for charge pumping induced by adiabatic tuning of $\gamma$. The dashed curve and solid curve are illustrations of charge density profile. By tuning $\gamma$ from $\infty$ to $0$, the charge is pumped in the direction of magnetic field $\vec{B}$ through the surface states. }
\end{figure*}

By including a new adiabatic parameter to the model, the chiral topological model can be used to describe Yang's monopole:
\begin{align}
        \tilde{H}=H+\gamma \sigma_z\otimes \mathbb{I}.
\end{align}
The adiabatic tuning of $\gamma$ from $\infty$ to $0$ induces charge pumping under the existence of magnetic field in the aligning direction; an illustration of the transport is shown in Fig.\ref{fig:CSchargePump}. Such transpot is usually described as TME.
The transport coefficient of charge pumping is related to the non-Abelian CS invariant~\cite{PhysRevB.78.195424},
\begin{align}
        \langle x^\mu\rangle|_{\gamma=0} - \langle x^\mu\rangle|_{\gamma=\infty} = I_{\text{CS}}[A_{\mu}]|_{\gamma=0} B^\mu,
\end{align}
where $B^{\mu}$ is the strength of magnetic field along $\mu$ direction and the non-Abelian CS invariant with $\gamma=0$ is
\begin{align}
        I_{\text{CS}}[A_{\mu}]=\frac{1}{ 2}\int d^3k~\text{Tr}[A_\mu\partial_\nu A_\rho +\frac{2}{ 3}A_\mu A_\nu A_\rho]\epsilon^{\mu\nu\rho},\label{eq:nonabelCS}
\end{align}
which relies on the non-Abelian Berry connection
\begin{align}
        A_\mu = \left[\begin{array}{cc}
                A_{\uparrow\uparrow,\mu}&A_{\uparrow\downarrow,\mu}\\
                A_{\downarrow\uparrow,\mu}&A_{\downarrow\downarrow,\mu}
        \end{array}\right],\quad \text{for }\mu=q_x,~q_y~\text{or}~q_z. \label{eq:nonabelA}
\end{align}
Each components of $A_\mu$ can be measured with different initial states $|x\pm\rangle = (1/\sqrt{2})(|\uparrow\rangle \pm |\downarrow\rangle)$ and $|y\pm\rangle = (1/\sqrt{2})(|\uparrow\rangle \pm i|\downarrow\rangle)$: 
\begin{align}
        &A_{x\pm,\mu}=\frac{i}{2\pi}\langle x\pm| \exp(ik_z\vec{h}(k_x,k_y)\cdot\vec{\sigma})\partial_\mu\exp(-ik_z\vec{h}(k_x,k_y)\cdot\vec{\sigma}) |x\pm\rangle=\frac{1}{2}(A_{\uparrow\uparrow,\mu}+A_{\downarrow\downarrow,\mu}\pm A_{\uparrow\downarrow,\mu}\pm A_{\downarrow\uparrow,\mu}),\\
        &A_{y\pm,\mu}=\frac{i}{2\pi}\langle y\pm| \exp(ik_z\vec{h}(k_x,k_y)\cdot\vec{\sigma})\partial_\mu\exp(-ik_z\vec{h}(k_x,k_y)\cdot\vec{\sigma}) |y\pm\rangle=\frac{1}{2}(A_{\uparrow\uparrow,\mu}+A_{\downarrow\downarrow,\mu}\pm i A_{\uparrow\downarrow,\mu}\mp i A_{\downarrow\uparrow,\mu}).
\end{align}
The definition of Eq. (\ref{eq:AdefBerryConn}) implies $A_{\uparrow \uparrow}=-A_{\downarrow\downarrow}$, so $A_{x+}=-A_{x-}$ and $A_{y+}=-A_{y-}$.
Then, $A_{\uparrow\downarrow,\mu}=2(A_{x+,\mu}-iA_{y+,\mu})$ and $A_{\downarrow\uparrow,\mu}=2(A_{x+,\mu}+iA_{y+,\mu})$.
Thus, we need to prepare three initial states $|\uparrow\rangle$, $|x+\rangle$ and $|y+\rangle$ followed by quantum quenches to get all the matrix elements of the non-Abelian Berry connection $A_{\mu}$.
However, as the Berry connection is not a physical quantity to be measured, a proper gauge is required to fix the Berry connection from the observables.
For a $SU(2)$ gauge, 3 gauge conditions in total are needed to fully fix the gauge freedom. We take the Coulumb gauge condition for the abelian gauge we measured, the gauge conditions and Berry curvature equations are as follows,
\begin{align}
        &\partial^\mu A_{s,\mu}=0,\quad s=\uparrow,~ x+~\text{or}~y+,\label{eq:gaugeCondNonAbel}\\
        &J_\mu=\epsilon_{\mu\nu\rho}\partial_\nu A_{s,\rho},\quad s=\uparrow,~x+~\text{or}~ y+.\label{eq:BerryConnBerryCurvNonAbel}
\end{align}

The observables are the expectation value of Pauli matrices with each momentum and each state of initial spin,
\begin{align}
        {P}_{s,\mu}(\bm{k})=\langle s|\exp(iq_z\bm{h}(q_x,q_y)\cdot\bm{\sigma})\sigma_{\mu}\exp(-iq_z\bm{h}(q_x,q_y)\cdot\bm{\sigma})|s\rangle\quad \text{for}~s=\uparrow,~x+~\text{or}~y+,
\end{align}
where $\bm{k}=q_x,q_y,q_z$.
Then the Berry curvature for different initial spins are
\begin{align}
        J_{s,\mu}(\bm{k})=\epsilon_{\mu\nu\rho}\epsilon_{abc}P_{s,a}(\bm{k})\partial_\nu P_{s,b}(\bm{k}) \partial_\rho P_{s,c}(\bm{k})/8\pi,\quad \text{for} ~s=\uparrow,~x+\text{or}~y+,
\end{align}
and the Berry connection can be obtained from Berry curvature
\begin{align}
{A}_{s,\mu}(\bm{k})=\int {J}_{s,\mu}(\bm{k}^{\prime})\times (\bm{k}-\bm{k}^{\prime})/|\bm{k}-\bm{k}^{\prime}|^3d\bm{k}^{\prime}/(4\pi),\quad \text{for} ~s=\uparrow,~x+\text{or}~y+.
\end{align}

All the components of the non-Abelian Berry connection Eq.~(\ref{eq:nonabelA}) can be evaluated through experimental observables ${P}_{s,\mu}(\bm{k})$, which renders an experimental probe of the the non-Abelian CS invariant according to Eq.~(\ref{eq:nonabelCS}).

\subsubsection{The experimental proposal to realize a 3D time-reversal-invariant topological insulator with non-Abelian CS invariant using Raman lattices}
The Chern-Simons invariant observed in the current work deals with two bands, which is a integral of the product of the \emph{Abelian} Berry curvature $\mathbf{J}(\mathbf{k})$ and the \emph{Abelian} Berry connection $\mathbf{A}(\mathbf{k})$.  As illustrated in the subsection above, it has a natural \emph{non-Abelian} generalization, $I_{\text{CS}}[{\cal{A}}]=\frac{1}{2}\int_{\text{BZ}}{\text{Tr}}({\cal{A}}d{\cal{A}}+\frac{2}{3}{\cal{A}}^3)$, where ${\cal{A}}^{\alpha\beta}({\mathbf{k}})= \frac{i}{2\pi} \left\langle u^\alpha({\mathbf{k}}) \right|\left.d u^\beta({\mathbf{k}})\right\rangle $ is the non-Abelian Berry connection. This non-Abelian CS invariant is just the strong invariant that classifies the 3D time-reversal-invariant topological insulators in class AII~\cite{Chiu2016}, which leads to the topological magneto-electric effect as a low-energy simulation of axion electrodynamics~\cite{Wilczek-axion-electrodynamics-1987,PhysRevB.78.195424}. We know show briefly how such 3D time-reversal-invariant topological insulators with nontrivial non-Abelian CS invariant could be engineered in our Raman lattice platform; more details will be elaborated in our subsequent work.

To have a nontrivial non-Abelian CS invariant, one needs at least four energy bands. The minimal model of the 3D time-reversal-invariant topological insulator is the 3D Bernevig-Hughes-Zhang (BHZ) model, which is an extension of the 2D BHZ model ~\cite{bernevig2006quantum} to the 3D case~\cite{zhang2009topological}. We consider the 3D BHZ model with the following form:
\begin{equation}\label{Eq:H_BHZ}
H_{\text{BHZ}}(\mathbf{q}) = \left(
\begin{array}{cc}
 H_\text{Weyl}(\mathbf{q}) & V_{eg}(\mathbf{q}) \\
 V_{eg}(\mathbf{q}) & H_\text{Weyl}^*(\mathbf{-q}) \\
\end{array}
\right),
\end{equation}
where $H_\text{Weyl}(\mathbf{q}) = \mathbf{h}(\mathbf{q})\cdot\bm{\sigma} = h_1(\mathbf{q})\sigma_x + h_2(\mathbf{q})\sigma_y + h_4(\mathbf{q})\sigma_z$ with $h_1(\mathbf{q}) = 2t_{\text{so}}\sin q_u$, $h_2(\mathbf{q}) = 2t_{\text{so}}\sin q_v$, $h_4(\mathbf{q}) = m_z - 2t_z \cos q_z - 2t_1(\cos q_u+\cos q_v)$ is the tight-binding model of Weyl semimetals simulated in our 3D Raman lattice platform~\cite{wang2021realization}, and $V_{eg}(\mathbf{q}) = \lambda \sin q_z \sigma_x \equiv h_3(\mathbf{q})\sigma_x$. The bottom-right block $H_\text{Weyl}^*(\mathbf{-q})  = \mathbf{h}'(\mathbf{q})\cdot\bm{\sigma} = -h_1(\mathbf{q})\sigma_x + h_2(\mathbf{q})\sigma_y + h_4(\mathbf{q})\sigma_z$ is the time-reversal conjugate of $H_\text{Weyl}(\mathbf{q})$. In the expression of $H_\text{Weyl}(\mathbf{q}) $, we have made use of a coordinate frame $(u,v)$ rotated from $(x,y)$ as $u = (x+y)/\sqrt{2}$ and $v = (x-y)/\sqrt{2}$, and $t_{\text{so}}$, $m_z$, $t_z$, $t_1$, $\lambda$ are all tunable parameters~\cite{wang2021realization}. By making use of four mutually anti-commute Dirac matrices $\Gamma_i$ ($i = 1,2, 3, 4$), we can rewrite $H_{\text{BHZ}}(\mathbf{q})$ in Eq.~(\ref{Eq:H_BHZ}) as follows:
\begin{equation}
H_{\text{BHZ}}(\mathbf{q}) = h_1(\mathbf{q})\Gamma_1 + h_2(\mathbf{q})\Gamma_2 + h_3(\mathbf{q})\Gamma_3 + h_4(\mathbf{q})\Gamma_4,
\end{equation}
where the Dirac matrices take the form $\Gamma_1 = s_z\otimes\sigma_x$, $\Gamma_2 = -s_0\otimes\sigma_y$, $\Gamma_3 = s_x\otimes\sigma_x$ and $\Gamma_4 = s_0\otimes\sigma_z$ with $s = (g, e)$ for $\lq\lq$spin" degree-of-freedom and $\sigma = (\uparrow, \downarrow)$ for $\lq\lq$orbital" degree-of-freedom. With the current choice of Dirac matrices, the time reversal operator that commutes with all the $\Gamma_i$'s takes the form ${\cal{T}} = i (s_y\otimes \sigma_z)K$, where $K$ is the complex conjugate operator. This leads to the doubly two-fold degenerate spectra of the four energy bands $E_{\pm}(\mathbf{q}) = \pm\sqrt{h_1(\mathbf{q})^2 + h_2(\mathbf{q})^2 + h_3(\mathbf{q})^2 + h_4(\mathbf{q})^2}$, which then yields the non-Abelian Berry curvature in the lowest two degenerate bands both with energy $E_-(\mathbf{q})$.

\begin{figure}
\begin{center}
\includegraphics[width=0.5\linewidth]{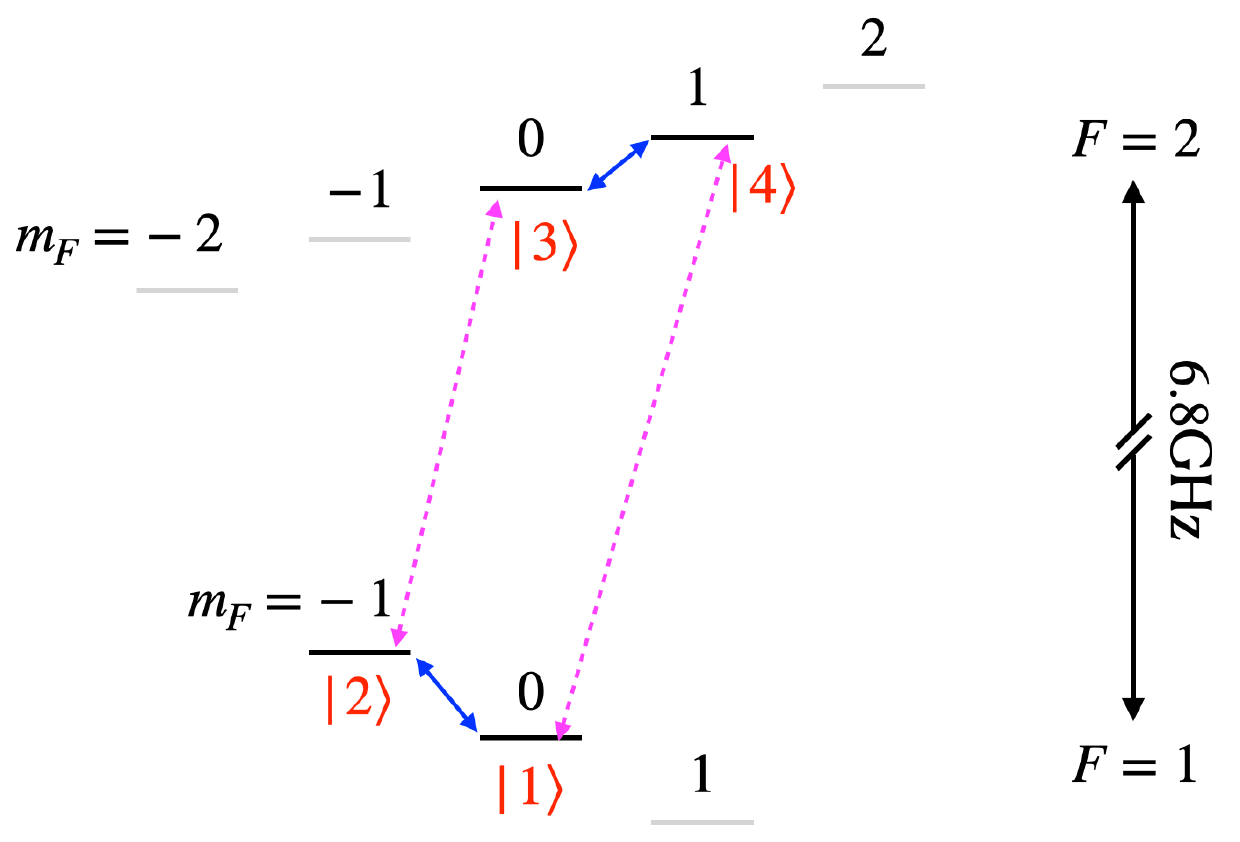}
\caption{Schematic of the proposed cyclic couplings using four hyperfine ground states of ${}^{87}$Rb. All the couplings are implemented by different configurations of Raman lasers, forming demanded \emph{Raman lattices} as described in the texts.
}
\label{Fig3DTI}
\end{center}
\end{figure}

To implement this four-band lattice model with cold atoms, we make use of cyclic couplings between four hyper-fine energy levels, which is similar to the scheme of Ref.~\cite{Second_C} but with a vital difference that the couplings in Ref.~\cite{Second_C} are spatially uniform using rf or microwave fields, while the couplings in our current proposal are spatially dependent using Raman lasers; see Fig.~\ref{Fig3DTI}. We choose four hyperfine states in the $\left| F, m_F \right\rangle$ and $\left| F'= F+1, m_{F'} \right\rangle$ manifolds; it applies to both bosonic species with integer-valued total spin $F$ and fermionic species with half-integer-valued $F$. In Fig.~\ref{Fig3DTI}, we consider the case of ${}^{87}$Rb, and denote $\left| F = 1, m_F = 0 \right\rangle = \left| g \uparrow \right\rangle \equiv \left| 1 \right\rangle$, $\left| F = 1, m_F = -1 \right\rangle = \left| g \downarrow \right\rangle \equiv \left| 2 \right\rangle$, $\left| F = 2, m_F = 0 \right\rangle = \left| e \uparrow \right\rangle \equiv \left| 3 \right\rangle$, and $\left| F = 2, m_F = 1 \right\rangle = \left| e \downarrow \right\rangle \equiv \left| 4 \right\rangle$. Then, we use Raman lasers to implement cyclic couplings between four hyper-fine energy levels: $\left| 1 \right\rangle \leftrightarrow \left| 2 \right\rangle$, $\left| 2 \right\rangle \leftrightarrow \left| 3 \right\rangle$, $\left| 3 \right\rangle \leftrightarrow \left| 4 \right\rangle$ and $\left| 4 \right\rangle \leftrightarrow \left| 1 \right\rangle$. To be concrete, the Raman coupling $\left| 1 \right\rangle \leftrightarrow \left| 2 \right\rangle$ could be taken as the form of the Weyl Hamiltonian, which implements $H_\text{Weyl}(\mathbf{q})$ as in our previous experiment in Ref.~\cite{wang2021realization}. The same sets of Raman lasers to implement $H_\text{Weyl}(\mathbf{q})$ could be frequency shifted via acousto-optic modulators and phase modulated via wave plates to implement the coupling $\left| 3 \right\rangle \leftrightarrow \left| 4 \right\rangle$, yielding another Weyl Hamiltonian $H_\text{Weyl}^*(\mathbf{-q})$.
In this way, the diagonal blocks of $H_{\text{BHZ}}(\mathbf{q})$ in Eq.~(\ref{Eq:H_BHZ}), $H_{\text{diag}}(\mathbf{q}) = H_\text{Weyl}(\mathbf{q})\oplus H_\text{Weyl}^*(\mathbf{-q}) = h_1(\mathbf{q})\Gamma_1 + h_2(\mathbf{q})\Gamma_2 + h_4(\mathbf{q})\Gamma_4$ is implemented using Raman lattices.
 We then align the Raman lasers along the $z$-direction to implement a one-dimensional Raman coupling (see, eg.~\cite{song2018observation}) $\left| 4 \right\rangle \leftrightarrow \left| 1 \right\rangle$ as well as $\left| 2 \right\rangle \leftrightarrow \left| 3 \right\rangle$, realizing the demanded off-diagonal blocks of $H_{\text{BHZ}}(\mathbf{q})$ in Eq.~(\ref{Eq:H_BHZ}), $ H_{\text{off-diag}}(\mathbf{q}) =  \lambda \sin q_z (\left| 1 \right\rangle \left\langle 4 \right| + \left| 2 \right\rangle \left\langle 3 \right| + \text{h.c.}) = h_3(\mathbf{q})\Gamma_3$. Now we have the building blocks $H_{\text{diag}}(\mathbf{q})$ (using the Raman coupling scheme in \cite{wang2021realization} as denoted by the blue double arrows in Fig.~\ref{Fig3DTI}) and $H_{\text{off-diag}}(\mathbf{q})$ (using the Raman coupling scheme in \cite{song2018observation} as denoted by the magenta dashed double arrows in Fig.~\ref{Fig3DTI}), it is straightforward to engineer a time-periodic Hamiltonian $\tilde H (\mathbf{q}, t)=\tilde H (\mathbf{q}, t+T)$ with periodicity $T$ as follows:
 \begin{equation}
\tilde H (\mathbf{q}, t) =
\begin{array}{cc}
 \left\{
\begin{array}{cc}
 2 H_{\text{diag}}(\mathbf{q}), &\text{for } 0\le t<T/2; \\
 2 H_{\text{off-diag}}(\mathbf{q}), &\text{for } T/2\le t<T. \\
\end{array}
\right.
 \\
\end{array}
 \end{equation}
 In the lowest order of $T$, the effective (Floquet) Hamiltonian of this periodically driven system $\tilde H (\mathbf{q}, t)$ is just the time-averaged Hamiltonian~\cite{Yu2016} $H_{\text{eff}}(\mathbf{q}) = H_{\text{diag}}(\mathbf{q}) + H_{\text{off-diag}}(\mathbf{q}) = H_{\text{BHZ}}(\mathbf{q})$.

\begin{thebibliography}{39}
\bibitem[{not({\natexlab{a}})}]{note3}
\bibinfo{note}{The Berry connection defined in this form is an Abelian one and
  applies to the case that the energy bands are non-degenerate; it leads to an
  Abelian CS invariant as in Eq.~(\ref{CSinvarint}). This Berry connection has
  a natural non-Abelian generalisation for the case that the energy bands are
  degenerate, which leads to an non-Abelian CS invariant.}
\bibitem[{not({\natexlab{b}})}]{note2}
\bibinfo{note}{The Berry connection can be obtained by the Berry curvature
  under a particular gauge}.
\end{thebibliography}
\end{document}